\documentclass{amsart}

\newtheorem{theorem}{Theorem}[section]

\theoremstyle{definition}
\newtheorem{definition}[theorem]{Definition}

\theoremstyle{remark}
\newtheorem{remark}[theorem]{Remark}
\numberwithin{equation}{section}
\newcommand{\abs}[1]{\lvert#1\rvert}

\newcommand{\q}{\mathfrak q}

\begin{document}

\title{VECTOR COHERENT STATES ON CLIFFORD ALGEBRAS}
\author{K. Thirulogasanthar$^\dagger$ }
\address{Department of Mathematics and Statistics, Concordia
University, 7141 Sherbrooke Street West, Montreal, Quebec H4B 1R6,
Canada } \email{${\dagger}$ ksanthar@hotmail.com}
\author{A.L. Hohou\'eto$^\ddagger$}
\email{${\ddagger}$ al$\_$hohoueto@yahoo.fr}
\date{\today}

\begin{abstract}
The well-known canonical coherent states are expressed as an
infinite series in powers of a complex number $z$ and a positive
sequence of real numbers $\rho(m)=m!$. In this article, in analogy
with the canonical coherent states, we present a class of vector
coherent states by replacing the complex variable $z$ by a real
Clifford matrix. We also present another class of vector coherent
states by simultaneously replacing $z$ by a real Clifford matrix
and $\rho(m)$ by a real matrix. As examples, we present vector
coherent states on quaternions and octonions with their real
matrix representations.
\end{abstract}

\maketitle

\section{Introduction}

Coherent states (CS for short) form an overcomplete family of
vectors in a Hilbert space. Typically, CS are quantum states and
known to describe quantum phenomenon favorably. In particular, CS
provide a mathematical tool to bring a close connection between
classical and quantum formalisms. The conventional CS can be
defined in a number of ways \cite{AAG}. One way to define a set of
CS is as follows.
\begin{definition}
Let $\mathfrak{H}$ be a Hilbert space with an orthonormal basis
$\{\phi_m\}_{m=0}^\infty$ and let $\mathbb{C}$ be the complex
plane. For $z \in \mathcal{D}$, an open subset of $\mathbb{C}$,
the states
 \begin{equation}
\mid z \rangle = \mathcal{N}(|z|)^{-1/2} \sum_{m=0}^\infty
\frac{z^m}{\sqrt{\rho(m)}} \phi_m
 \label{1}
 \end{equation}
are said to form a set of CS if
 \begin{enumerate}
 \item[(a)] The states $\mid z \rangle$ are normalized, that is,
$\langle z \mid z \rangle = 1$,
 \item[(b)] The states $\mid z \rangle$ give a resolution of the
identity, that is,
 \begin{equation}
\int_{\mathcal{D}} \mid z \rangle W(|z|) \langle z \mid d\mu = I,
 \label{2}
 \end{equation}
 \end{enumerate}
where $\mathcal{N}(|z|)$ is a normalization factor,
$\{\rho(m)\}_{m=0}^\infty$ is a sequence of nonzero positive real
numbers,  $W(|z|)$ is a positive weight function, $d\mu$ is an
appropriately chosen measure on $\mathcal{D}$, and $I$ is the
identity operator on $\mathfrak{H}$.
 \label{d1}
 \end{definition}
In \cite{KA} vector coherent states (VCS for short) were presented
in the form (\ref{1}) by replacing the complex number $z$ by an
$n\times n$ matrix $Z = A(r) e^{i\zeta\Theta(k)}$ with the
conditions
 \begin{eqnarray*}
A(r)A(r)^\dagger &=& A(r)^\dagger A(r), \\
\Theta(k)^\dagger &=& \Theta(k), \quad \text{and} \quad
\Theta(k)^2 = \mathbb{I}_n \\
\left[ A(r), \Theta(k) \right] &=& 0
 \end{eqnarray*}
imposed on the $n \times n$ matrices $A(r)$ and $\Theta(k)$, where
the superscript $\dagger$ stands for the complex conjugate
transpose of a matrix, and $\mathbb{I}_n$ is the $n \times n$
identity matrix. As an example, VCS were constructed from the
complex representation of the quaternions. Further, these VCS,
stemming from the complex matrix representation of quaternions, were analized to a certain extent.

In this article, we present VCS on real Clifford algebras by
considering their real matrix representations. Here, we do not
consider the matrix in the form $Z = A(r) e^{i \zeta \Theta(k)}$
and we do not impose any conditions on the matrices, but, in
addition to the Clifford parameters, we introduce a new parameter
$\theta$ whose role in the construction will be made clear later.
\section{On a general Clifford algebra}
Let $X$ be a real linear space and $(X,f)$ be a real quadratic
space. By this, we mean that
 \begin{equation}
f = \langle \cdot \mid \cdot \rangle : X \times X \rightarrow
\mathbb{C}, \quad (x,y) \mapsto f(x,y)
 \end{equation}
is the usual inner product on $X$. Now, the Clifford algebra of
$(X,f)$ is a pair $(C(f),\Theta )$, where $C(f)$ is a
$\mathbb{R}$-algebra and
 \begin{equation}
\Theta : X \rightarrow C(f), \quad x \mapsto \Theta(x)
 \end{equation}
is a linear function such that
 \begin{equation}
\Theta(x) \Theta(x)^T = \Theta(x)^T \Theta(x) = f(x,x)
\mathbb{I}_n = \|x\|^2 \mathbb{I}_n, \quad \forall ~x \in X,
 \label{clif11}
 \end{equation}
where we have taken an $n \times n$ matrix representation of
$C(f)$ by real matrices satisfying (\ref{clif11}) and denoted
these matrices by the same symbol as the elements of the algebra
$C(f)$. For each $z \in S^1$, the unit circle, we define a new
linear function $\Theta_z(x) = z \Theta(x)$. The new function
satisfies the relation
 \begin{equation}
\Theta_z(x) \Theta_z(x)^\dagger = \Theta_z(x)^\dagger \Theta_z(x)
= \Theta(x)^T \Theta(x) = f(x,x) \mathbb{I}_n = \|x\|^2
\mathbb{I}_n \quad \forall ~x \in X.
 \label{clif1}
 \end{equation}
Therefore, for each $z\in S^1$, the pair $(C(f),\Theta_{z})$ is
again a Clifford algebra. Further, it can be noticed that, for any
two different $z\in S^1$, the corresponding Clifford algebras are
isomorphic. \\
Let $\chi^1, \ldots, \chi^n$ be an orthonormal basis of
$\mathbb{C}^n$, and let $\{\phi_m\}_{m=0}^\infty$ be an
orthonormal basis of an arbitrary Hilbert space $\mathfrak{H}$.
With these considerations, we define on $\mathbb{C}^n \otimes
\mathfrak{H}$ the states
 \begin{equation}
\mid \Theta_z,j \rangle = \mid \Theta_z \rangle^j =
\mathcal{N}^{-1/2}(\|x\|) \sum_{m=0}^\infty
\frac{\Theta_z(x)^m}{\sqrt{\rho(m)}} \chi^j \otimes \phi_m, \quad
j = 1, 2, \ldots, n,
 \label{clif2}
 \end{equation}
where the normalization constant $\mathcal{N}$ and $\rho(m)$ have
to be identified so that the states (\ref{clif2}) satisfy the
normalization condition and a resolution of the identity.

\subsection{Normalization and Resolution of the Identity}

From (\ref{clif1}) we have that
$$
(\Theta_z(x)^\dagger)^m \Theta_z(x)^m = \|x\|^{2m} \mathbb{I}_n.
$$
Let us make the following identification
 \begin{equation}
 \label{id}
\|\cdot\| : X \rightarrow \mathbb{R}_+, \qquad \|x\| \equiv t,
\end{equation}
and set
 \begin{equation}
x_m = \frac{\rho(m)}{\rho(m-1)}, \quad \forall ~m \geq 1.
 \end{equation}
We can then define
 \begin{eqnarray}
x_m! &=& \rho(m), \quad m \geq 1 \\
x_0! &=& 1. \nonumber
 \end{eqnarray}
Assuming that
 \begin{equation}
\lim_{m \to \infty} x_m = R,
 \end{equation}
the set of labels becomes
 \begin{equation}
\mathcal{D} = \{(t,\theta) \; | \; 0 \leq t < L, \; 0 \leq \theta
\leq 2\pi\}, \quad l = \sqrt{R},
 \end{equation}
with the measure $d\mu = t dt d\theta$, where $d\theta$ is a
measure on $S^1$, whose elements are parametrized as
$e^{i\theta}$, $\theta \in [0,2\pi]$. We can now state the
following
 \begin{theorem}
The states in (\ref{clif2}) are normalized in the sense that
 \begin{equation}
\sum_{j=1}^n \langle \Theta_z,j \mid \Theta_z,j \rangle = 1,
 \end{equation}
and they achieve a resolution of the identity
 \begin{equation}
\int_0^L \int_0^{2\pi} w(t) \sum_{j=1}^n \mid \Theta_z,j \rangle
\langle \Theta_z,j \mid t dt d\theta = \mathbb{I}_n \otimes I,
 \end{equation}
where
 \begin{eqnarray}
\mathcal{N}(t) &=& n \sum_{m=0}^\infty \frac{t^{2m}}{x_m!}, \\
w(t) &=& \frac{\mathcal{N}(t)}{2\pi} \lambda(t),
 \end{eqnarray}
and $\lambda(t)$ is to be chosen from the moment problem
 \begin{equation}
\int_0^L \lambda(t) t^{2m+1} dt = x_m!.
 \label{m1}
 \end{equation}
 \label{clift1}
 \end{theorem}
 \begin{proof}
It is  straightforward that, for $(t,\theta) \in \mathcal{D}$, the
normaliszation condition reads
 \begin{eqnarray*}
\sum_{j=1}^n \langle \Theta_z,j \mid \Theta_z,j \rangle &=&
\mathcal{N}(t)^{-1} \sum_{j=1}^n \sum_{m=0}^\infty
\sum_{l=0}^\infty \frac{1}{\sqrt{\rho(m) \rho(l)}} \langle
\Theta_z(x)^{l\dagger} \Theta_z(x)^m \chi^j \mid \chi^j \rangle
\langle \phi_m \mid \phi_l \rangle \\
&=& \mathcal{N}(t)^{-1} \sum_{j=1}^n \sum_{m=0}^\infty
\frac{1}{\rho(m)} t^{2m} \langle \chi^j \mid \chi^j \rangle =
n \mathcal{N}(t)^{-1} \sum_{m=0}^\infty \frac{t^{2m}}{\rho(m)} \\
&=& 1,
 \end{eqnarray*}
that is,
$$
\mathcal{N}(t) = n \sum_{m=0}^\infty \frac{t^{2m}}{x_{m}!}.
$$
In the case, $x_m=m$, for all $m$, $\mathcal{N}(t) = n e^{t^2}$.
\newline
Let us turn now to the condition leading to a resolution of the
identity. We have that
 \begin{eqnarray*}
\lefteqn{ \int_0^L \int_0^{2\pi} w(t) \sum_{j=1}^n \mid \Theta_z,j
\rangle \langle \Theta_z,j \mid t dt d\theta = {} } \\
 &=& {} \sum_{j=1}^n \sum_{m=0}^\infty \sum_{l=0}^\infty \int_0^L
\int_0^{2\pi} \frac{w(t)}{\mathcal{N}(t) \sqrt{\rho(m) \rho(l)}}
\mid \Theta_z^m \chi^j \otimes \phi_m \rangle
\langle \Theta_z^l \chi^j \otimes \phi_l \mid d\mu {} \\
&=& {} \sum_{j=1}^n \sum_{m=0}^\infty \sum_{l=0}^\infty \int_0^L
\int_0^{2\pi} \frac{w(t)}{\mathcal{N}(t) \sqrt{\rho(m) \rho(l)}}
\Theta_z^m \mid \chi^j \rangle \langle \chi^j \mid
\Theta_z^{l\dagger} \otimes \mid \phi_m \rangle \langle \phi_l \mid
d\mu {} \\
&=& {} \sum_{m=0}^\infty \sum_{l=0}^\infty \int_0^L \int_0^{2\pi}
\frac{w(t)}{\mathcal{N}(t) \sqrt{\rho(m) \rho(l)}} \Theta_z^{m}
\mathbb{I}_n \Theta_z^{l\dagger} \otimes \mid \phi_m \rangle
\langle \phi_l \mid d\mu {} \\
&=& {} \sum_{m=0}^\infty \sum_{l=0}^\infty \int_0^L \int_0^{2\pi}
\frac{w(t)}{\mathcal{N}(t) \sqrt{\rho(m) \rho(l)}}
e^{i(m-l)\theta} \Theta(x)^m \Theta(x)^{l\dagger} \otimes
\mid \phi_m \rangle \langle \phi_l \mid d\mu {} \\
&=& {} \sum_{m=0}^\infty \int_0^L \frac{2\pi w(t)}{\mathcal{N}(t)
\rho(m)} \Theta(x)^m \Theta(x)^{m\dagger} \otimes \mid \phi_m \rangle
\langle \phi_m \mid t dt {} \\
&=& {} \sum_{m=0}^\infty \int_0^L \frac{2\pi w(t)}{\mathcal{N}(t)
x_m!} t^{2m} t dt \mathbb{I}_n \otimes \mid \phi_m \rangle
\langle \phi_m \mid {} \\
 \end{eqnarray*}
choosing $\displaystyle w(t) = \frac{\mathcal{N}(t)}{2\pi}
\lambda(t)$, we obtain that
 \begin{eqnarray*}
\lefteqn{ \int_0^L \int_0^{2\pi} w(t) \sum_{j=1}^n \mid \Theta_z,j
\rangle \langle \Theta_z,j \mid t dt d\theta = {} } \\
&=& {} \sum_{m=0}^\infty \frac{1}{x_m!} \left[ \int_0^L \lambda(t)
t^{2m+1} dt \right] \mathbb I_n \otimes \mid \phi_m \rangle
\langle \phi_m \mid {} \\
&=& {} \sum_{m=0}^\infty \mathbb I_n \otimes \mid \phi_m \rangle
\langle \phi_m \mid {} \\
&=& {} \mathbb I_n \otimes I, {}
 \end{eqnarray*}
provided the function $\lambda$ is such that
$$
\int_0^L \lambda(t) t^{2m+1} dt = x_m!.
$$
This condition is achievable: For example, when $x_m=m$ (or
$\rho(m)=m!$), we have $L=\infty$, $w(t) = n/\pi$, and the choice
$\lambda(t)=e^{-t^2}$, together with the substitution $t^2=r$,
give the expected resolution of the identity through the equation
$$
\int_{0}^\infty e^{-r} r^{(m+1)-1} dr = \Gamma(m+1).
$$
 \end{proof}
Theorem (\ref{clift1}) shows that the set of states in
(\ref{clif2}) are VCS.

In the above construction we have introduced an additional complex
number $e^{i\theta}$ from $S^1$ to make our calculations easier.
It can easily be noted that the new parameter intervenes in the
picture only by  bringing the double sum in the resolution of the
identity condition to a single sum through the identity
$$
\int_0^{2\pi} e^{i(m-l)\theta} d\theta =
\left\{
\begin{array}{ccc}
0 & \text{if}& m \neq l \\ 2 \pi & \text{if} & m = l
\end{array}
\right. .
$$
Since the moment problem (\ref{m1}) is in the form of a classical
moment problems, $\rho(m)$ can be chosen in several ways. For
example, several moment problems were solved in \cite{KPS}.

\section{Annihilation, Creation, and Number operators}

Here, we briefly show that the approach adopted in \cite{KA} for
the complex representation of quaternions can be applied to any
Clifford algebra. We define the annihilation, creation and the
number operator on the basis $\{\phi_m\}$ by
 \begin{eqnarray}
a \phi_m &=& \sqrt{x_m} \phi_{m-1} \\
a^\dagger \phi_m &=& \sqrt{x_{m+1}} \phi_{m+1}\\
N' \phi_m &=& x_m \phi_m.
 \end{eqnarray}
The corresponding operators for the states $\mid \Theta_z(x),j
\rangle$ are
 \begin{eqnarray}
A &=& \mathbb I_n \otimes a \\
A^\dagger &=& \mathbb I_n \otimes a^\dagger \\
N &=& \mathbb I_n \otimes N'
 \end{eqnarray}
The action of these operators is given by the same relations of
\cite{KA}. Further, when $x_m=m$ we have
 \begin{equation}
\mid \Theta_z(x),j \rangle = \frac{1}{\sqrt{n}} e^{\Theta_z(x)
\otimes a^\dagger - \Theta_z(x)^\dagger \otimes a} \chi^j \otimes
\phi_0.
 \label{e1}
 \end{equation}
The proof is similar to the proof presented in \cite{KA}.

Let us define the self-adjoint operators
 \begin{equation}
\widehat{q} = \frac{1}{\sqrt{2}} (a+a^\dagger) \quad,\quad
\widehat{p} = \frac{1}{i\sqrt{2}} (a-a^\dagger),
 \end{equation}
and the corresponding operators for the VCS as
 \begin{equation}
Q = \frac{1}{\sqrt{2}} (A+A^\dagger) \quad,\quad P =
\frac{1}{i\sqrt{2}} (A-A^\dagger).
 \end{equation}
When $\rho(m)=m!$, the CS in (\ref{1}) can be written as
 \begin{equation}
\mid z \rangle = e^{z a^\dagger - \overline{z} a} \phi_0 = e^{i p
\widehat{q} - q \widehat{p}} \phi_0, \quad z = \frac{1}{\sqrt{2}}
(q-ip).
 \end{equation}
The operators $\widehat{q}$, $\widehat{p}$ and $I$ generate an
irreducible representation of the Lie algebra $\mathfrak{g}_{WH}$
of the Weyl-Heisenberg group $G_{WH}$ on the Hilbert space
$\mathfrak{H}$. A unitary irreducible representation of $G_{WH}$
on $\mathfrak{H}$ is given by the operators $U(\theta,q,p) =
e^{i(\theta I + p \widehat{p} + q \widehat{q})}$. Thus $\mid z
\rangle = U(0,q,p) \phi_0$. As it was given in \cite{KA} an exact
anologue follows for the VCS, $\mid \Theta_z(x),j \rangle$. The
operators $A$, $A^\dagger$, $N$ generate an algebra (under the
commutator bracket) $\mathfrak{U}_{\text{osc}}$, the so-called
oscillator algebra. The nature of this algebra primarily depends
on the choice of $\rho(m)$.

\section{Examples}

\subsection{Example 1 : Quaternionic VCS}
\label{qua}

Here, we present quaternionic VCS with the real matrix
representation of quaternions.

Let
 \begin{equation}
\mathbb{H} = \{\q' = a_0 + a_1i + a_2j + a_3k \; \mid \; i^2 = j^2
= k^2 = -1, \; ijk =-1, \; a_0,a_1,a_2,a_3 \in \mathbb{R}\}
 \end{equation}
be the real quaternion division algebra. It is known that
$\mathbb{H}$ is algebraically isomorphic to the real matrix
algebra
 \begin{equation}
\mathcal{M} = \left\{ \q'=
 \left(
\begin{array}{cccc}
a_0 & -a_1 & -a_2 & -a_3 \\
a_1 & a_0 & -a_3 & a_2 \\
a_2 & a_3 & a_0 & -a_1 \\
a_3 & -a_2 & a_1 & a_0
\end{array}
 \right)
\; | \; a_0,a_1,a_2,a_3 \in \mathbb{R} \right\}.
 \end{equation}
For detailed explanation see \cite{T} and the references therein.

For $z=e^{i\theta} \in S^1$ and $\q' \in \mathcal{M}$, let $\q = z
\q'$. Then,
$$
\q \q^\dagger = \q^\dagger \q = (a_0^2 + a_1^2 + a_2^2 + a_3^2)
\mathbb{I}_4 = |\q|^2 \mathbb{I}_4 = |\q'|^2 \mathbb{I}_4,
$$
where $|\q'|$ is the norm of the quaternion $\q'$ and
$\mathbb{I}_4$ is the $4 \times 4$ identity matrix. Thus, with the
notations of the previous sections we have a set of VCS in the
Hilbert space $\mathbb{C}^4 \otimes \mathfrak{H}$,
 \begin{equation}
\mid \q,j \rangle = \frac{1}{2} e^{-|\q|^2/2} \sum_{m=0}^\infty
\frac{\q^m}{\sqrt{m!}} \chi^j \otimes \phi_m, \quad j=1,\ldots,4.
 \end{equation}
Here also, the choice $\rho(m)=m!$ can be replaced by other
choices as mensioned earlier, because the moment condition takes
the form (\ref{m1}).

\subsubsection{Quaternionic minimum uncertainty states}

The eigenvalues of $\q$ are $z_1=a_0+ib$ and $z_2=a_0-ib$, each
with multiplicity 2, where $b = \sqrt{a_1^2+a_2^2+a_3^2}$. Let the
orthonormal eigenvectors (they do exist) corresponding to $z_1$
and $z_2$ be $\chi^{(1)}_j$, $\chi^{(2)}_j$, $j=1,2$ respectively.
Since
$$
\q^m \chi^{(1)}_j = z_1^m \chi^{(1)}_j, \quad \text{and} \quad
\q^m \chi^{(2)}_j = z_2^m \chi^{(2)}_j,
$$
the states
 \begin{equation}
\mid \q,j,i \rangle = \frac{1}{2} e^{-|\q|^2/2} \sum_{m=0}^\infty
\frac{z_i^m}{\sqrt{m!}} \chi^{(i)}_j \otimes \phi_m, \quad j,i =
1,2
 \end{equation}
saturate the Heisenberg uncertainty relation
 \begin{equation}
\langle \Delta Q_L \rangle \langle \Delta P_L \rangle \geq
\frac{1}{2}.
 \label{hur}
 \end{equation}

\subsubsection{The exponential form}

As introduced in the general case, we can take the annihilation,
creation, and number operators for the VCS as
 \begin{equation}
A = \mathbb{I}_4 \otimes a \quad,\quad A^\dagger =\mathbb{I}_4
\otimes a^\dagger \quad,\quad N = \mathbb{I}_4 \otimes N'.
 \end{equation}
With these operators we can write the VCS in the form
 \begin{equation}
\mid \q,j \rangle = \frac{1}{2} e^{\q \otimes a^\dagger -
\q^\dagger \otimes a} \chi^j \otimes \phi_0, \quad j=1,\ldots,4.
 \end{equation}
Once again, the proof is similar to the one presented in
\cite{KA}.

\subsection{Example 2 : Octonionic VCS}
\label{oct}

Let $\mathbb{O}$ denotes the octonion algebra over the real number
field $\mathbb{R}$. In \cite{T} it was shown that any $a \in
\mathbb{O}$ has a left matrix representation $\omega(a)$ and a
right matrix representation $\nu(a)$, given respectively by
 \begin{equation}
\omega(a) = \left(
\begin{array}{cccccccc}
a_0 & -a_1 & -a_2 & -a_3 & -a_4 & -a_5 & -a_6 & -a_7 \\
a_1 & a_0 & -a_3 & a_2 & -a_5 & a_4 & a_7 & -a_6 \\
a_2 & a_3 & a_0 & -a_1 & -a_6 & -a_7 & a_4 & a_5 \\
a_3 & -a_2 & a_1 & a_0 & -a_7 & a_6 & -a_5 & a_4 \\
a_4 & a_5 & a_6 & a_7 & a_0 & -a_1 & -a_2 & -a_3 \\
a_5 & -a_4 & a_7 & -a_6 & a_1 & a_0 & a_3 & -a_2 \\
a_6 & -a_7 & -a_4 & a_5 & a_2 & -a_3 & a_0 & a_1 \\
a_7 & a_6 & -a_5 & -a_4 & a_3 & a_2 & -a_1 & a_0
\end{array}
\right),
 \end{equation}
and
 \begin{equation}
\nu(a) = \left(
\begin{array}{cccccccc}
a_0 & -a_1 & -a_2 & -a_3 & -a_4 & -a_5 & -a_6 & -a_7 \\
a_1 & a_0 & a_3 & -a_2 & a_5 & -a_4 & -a_7 & a_6 \\
a_2 & -a_3 & a_0 & a_1 & a_6 & a_7 & -a_4 & -a_5 \\
a_3 & a_2 & -a_1 & a_0 & a_7 & -a_6 & a_5 & -a_4 \\
a_4 & -a_5 & -a_6 & -a_7 & a_0 & a_1 & a_2 & a_3 \\
a_5 & a_4 & -a_7 & a_6 & -a_1 & a_0 & -a_3 & a_2 \\
a_6 & a_7 & a_4 & -a_5 & -a_2 & a_3 & a_0 & -a_1 \\
a_7 & -a_6 & a_5 & a_4 & -a_3 & -a_2 & a_1 & a_0
\end{array}
\right).
\end{equation}
The relationship between the two representations is given by the
equation
 \begin{equation}
\nu(a) = K_8 \omega(a)^T K_8,
 \label{link-nu-omega}
 \end{equation}
where $K_8 = \text{diag}(K_4, \mathbb{I}_4)$ is an orthogonal
matrix, $K_4 = \text{diag}(1,-1,-1,-1)$ being the metric of the
space of Minkowski $\mathbb{R}_{1,3}^4$.

Let $z = e^{i\theta} \in S^1$, and define
 \begin{equation}
\omega(a,z) = z \omega(a), \quad \text{and} \quad \nu(a,z) = z
\nu(a).
 \end{equation}
Then,
\begin{eqnarray*}
\omega(a,z) \omega(a,z)^\dagger &=& \omega(a,z)^\dagger
\omega(a,z) = \nu(a,z) \nu(a,z)^\dagger = \nu(a,z)^\dagger \nu(a,z) \\
&=& (a_0^2 + a_1^2 + a_2^2 + a_3^2 + a_4^2 + a_5^2 + a_6^2 +
a_7^2) \mathbb{I}_8 \\
&=& \|a\|^2 \mathbb{I}_8.
\end{eqnarray*}
Thus, we have two sets of VCS
\begin{eqnarray}
\mid \omega(a,z),j \rangle &=& \frac{1}{\sqrt{8}} e^{-\|a\|^2/2}
\sum_{m=0}^\infty \frac{\omega(a,z)^m}{\sqrt{m!}}
\chi^j \otimes \phi_m, \quad j = 1,\ldots,8 \\
\mid \nu(a,z),j \rangle &=& \frac{1}{\sqrt{8}} e^{-\|a\|^2/2}
\sum_{m=0}^\infty \frac{\nu(a,z)^m}{\sqrt{m!}} \chi^j \otimes
\phi_m \quad j = 1,\ldots,8.
\end{eqnarray}
In fact, using(\ref{link-nu-omega}), the "right" VCS can be
obtained from the "left" ones by the transform
 \begin{equation}
\mid \nu(a,z),j \rangle = K_8 \mid \omega(\overline{a},z),j
\rangle K_8, \quad a \in \mathbb{O}, \; j = 1,\ldots,8.
 \end{equation}
Again, the moment condition takes the form (\ref{m1}). Thus,
$\rho(m)$ can be chosen in many ways. We keep stuck with the
simplest choice $\rho(m)=m!$.

\subsubsection{The oscillator algebra}

Now let us denote the corresponding annihilation, creation, and
number operators as before and define accordingly, for the left
representation,
 \begin{equation}
A_L, \; A_L^\dagger, \; N_L,
 \end{equation}
and, for the right representation,
 \begin{equation}
A_R, \; A_R^\dagger, \; N_R.
 \end{equation}
The action and commutation relations of these operators, and the
corresponding algebras take the same form as in the previous
sections. In principle, these two sets of operators generate two
algebras $\mathfrak{U}_{\text{osc}}^L$ and
$\mathfrak{U}_{\text{osc}}^R$, but they are the same.

\subsubsection{Octonionic minimum uncertainty states}

Let us denote the self-adjoint operators for the left
representation by $Q_L$ and $P_L$. Let
$$
b = \sqrt{a_1^2+a_2^2+\ldots+a_7^2}.
$$
Then, the eigenvalues of $\omega(a,z)$ are $z_1 =
e^{i\theta}(a_0+ib)$ and $z_2 = e^{i\theta}(a_0-ib)$, each with
multiplicity 4. Let the corresponding normalized eigenvectors be
$\chi^{(1)}_j$, $\chi^{(2)}_j$, $j=1,\ldots,4$. For $i=1,2$, the
states
 \begin{eqnarray}
\mid \omega(a,z),j,i \rangle &=& \frac{1}{\sqrt{8}} e^{-\|a\|^2/2}
\sum_{m=0}^\infty \frac{\omega(a,z)^m}{\sqrt{m!}}
\chi^{(i)}_j \otimes \phi_m \\
&=& \frac{1}{\sqrt{8}} e^{-\|a\|^2/2} \sum_{m=0}^\infty
\frac{z_i^m}{\sqrt{m!}} \chi^{(i)}_j \otimes \phi_m, \quad
j=1,\ldots,4 \nonumber
 \end{eqnarray}
saturate the Heisenberg uncertainty relation (\ref{hur}).

A similar set of VCS can also be obtained for the right
representation.

\subsubsection{The exponential form}

Here again the states can be written in the form
 \begin{equation}
\mid \omega(a,z),j \rangle = \frac{1}{\sqrt{8}} e^{\omega(a,z)
\otimes A_L^\dagger - \omega(a,z)^\dagger \otimes A_L} \chi^j
\otimes \phi_0,
 \end{equation}
and
 \begin{equation}
\mid \nu(a,z),j \rangle = \frac{1}{\sqrt{8}} e^{\nu(a,z) \otimes
A_R^\dagger - \nu(a,z)^\dagger \otimes A_R} \chi^j \otimes \phi_0.
  \end{equation}

\section{VCS with matrix moments}

So far, we have obtained VCS by replacing the complex number $z$
of (\ref{1}) by a Clifford matrix and keping the moments
$\rho(m)$, as usual, a positive sequence of real numbers. In this
section, we attempt to replace both, $z$ and $\rho(m)$ by
matrices, namely, the first by a Clifford matrix, and the second
by a matrix $R(m)$. This procedure foreshadows a more general
picture to be analyzed in a forthcoming paper.
 \newline
In order to do this we demand the matrix $R(m)$ to satisfy the
condition,
 \begin{equation}
R(m)R(m)^\dagger = R(m)^\dagger R(m) = f(m) \mathbb{I}_n,
 \label{m1}
 \end{equation}
where $n$ has to be chosen to match the size of the Clifford
matrix. Now, we define our VCS as follows:
 \begin{equation}
\mid Z,j \rangle = \mathcal{N}(\abs{Z})^{-1/2} \sum_{m=0}^\infty
R(m) Z^m \chi^j \otimes \phi_m, \quad j = 1,\ldots,n.
 \end{equation}
Let us then look at the normalization and resolution of the
identity conditions for the matrices of sections \ref{qua} and
\ref{oct}.

\subsection{For the quaternions}

Let $\q \in \mathbb{H}$, $Z = \q e^{i\theta}$, and, for $R(m)$,
let us take, for example, for a fixed $x$,
 \begin{equation}
R(m) = \frac{1}{\sqrt{m!}} \left(
 \begin{array}{cc}
\mathbb{I}_2 \cos{x} & -\mathbb{I}_2 \sin{x} \\
\mathbb{I}_2 \sin{x} & \mathbb{I}_2 \cos{x}
 \end{array}
\right).
 \end{equation}
We have
$$
R(m)R(m)^\dagger =R(m)^\dagger R(m) = \frac{1}{m!} \mathbb{I}_4.
$$
The normalization condition takes the form,
 \begin{eqnarray*}
\sum_{j=1}^4 \langle Z,j \mid Z,j \rangle &=&
\mathcal{N}(|Z|)^{-1} \sum_{j=1}^4 \sum_{m=0}^\infty \langle R(m)
\q^m \chi^j \mid R(m) \q^m \chi^j \rangle \\
&=& 4 \mathcal{N}(|Z|)^{-1} \sum_{m=0}^\infty \frac{|\q|^{2m}}{m!}
= 4 \mathcal{N}(|Z|)^{-1} e^{|\q|^2} =1,
 \end{eqnarray*}
and leads to the normalization factor
 \begin{equation}
\mathcal{N}(|Z|) = 4 e^{|\q|^2}.
 \end{equation}
On the other hand, under the identification (\ref{id}) and using
the measure
 \begin{equation}
d\mu = \frac{4|\q| d|\q| d\theta}{\pi},
 \end{equation}
we have
\begin{eqnarray*}
\lefteqn{ \sum_{j=1}^4 \int_0^\infty \int_0^{2\pi} \mid Z,j \rangle
\langle Z,j \mid d\mu = {} } \\
&=& {} \sum_{m=0}^\infty \sum_{l=0}^\infty \int_0^\infty
\int_0^{2\pi} e^{i(m-l)\theta} R(m) \q^m \left[ \sum_{j=0}^\infty
\mid \chi^j \rangle \langle \chi^j \mid \right] [R(l)
\q^l]^\dagger \otimes {} \\
& & {} \hspace{2cm} \otimes \mid \phi_m \rangle \langle \phi_l
\mid e^{-|\q|^2} \frac{|\q|d|\q|d\theta}{\pi} {} \\
&=& {} \sum_{m=0}^\infty \frac{1}{m!} \left[ \int_0^\infty
e^{-|\q|^2} |\q|^{2m} 2|\q|d|\q| \right] \mathbb{I}_4 \otimes \mid
\phi_m \rangle \langle \phi_l \mid {} \\
&=& {} \mathbb{I}_4 \otimes I. {}
\end{eqnarray*}

\subsection{For the octonions}

Here we can take either $Z = \omega(a,z)$ or $Z = \nu(a,z)$, where
$\omega(a,z)$ and $\nu(a,z)$ are as in section \ref{oct}. For
$R(m)$, one could take, for a fixed $x$,
 \begin{equation}
R(m) = \frac{1}{\sqrt{m!}} \left(
\begin{array}{cc}
\mathbb{I}_4 \cos{x} & -\mathbb{I}_4 \sin{x} \\
\mathbb{I}_4 \sin{x} & \mathbb{I}_4 \cos{x} \\
\end{array}
\right).
 \end{equation}
The rest of the details are similar to the case of quaternions.

 \begin{remark}
Since $Z$ and $R(m)$ are matrices, in general, placing $R(m)$ on
the right of $Z$ is different from placing it on the left of $Z$.
In our case, by the properties of the Clifford matrices and by the
assumption on $R(m)$ the construction can be carried out in either
way without any obstacles. Further, in CS constructions the
sequence $\rho(m)$ of (\ref{1}) is taken to be a positive
sequence, and, thereby, in getting a resolution of the identity,
we end up with a positive moment problem. When we replace
$\rho(m)$ by a matrix $R(m)$, one could expect it to be a positive
definite matrix. But in our construction we have used a
non-positive definite matrix $R(m)$. It can be observed that, it
doesn't really matter which matrix we start with, but what does
matter is that we finally end up with a moment problems with
positive moments throughout in the construction.
 \end{remark}

 \begin{remark}
In our knowledge, a physical system which can be described by
quaternions or octonions is not known yet. There have been several
attempts to describe relativistic physics in terms of quaternions
or octonions (see \cite{DD} and the references therein), but
without the expected success. Nevertheless, the VCS of the type
presented here have been introduced recently \cite{KA}, and
quaternionic VCS are suspected to be useful in the description of
the spin orbit interaction between a spinning electron and an
external magnetic field.
 \end{remark}

\section{conclusion}

We have presented a class of VCS using the real matrix
representation of Clifford algebras. In \cite{KA}, quaternionic
VCS were presented without introducing the supplementary term
$e^{i\theta}$, whis is used in the construction developed in this
paper only for technical purposes. Finally, we have introduced VCS
with matrix moments, carrying out the construction for real matrix
representations of quaternions and octonions. A wide-ranging study
of VCS with matrix moments will emerge in a companion paper.

\end{document}